# A Partial Order on Bipartite Graphs with *n* Vertices

**Emil Daniel Schwab**
**Department of Mathematical Sciences, University of Texas - El Paso, USA**

**ABSTRACT.** The paper examines a partial order on bipartite graphs $(X_1, X_2, E)$ with *n* vertices, $X_1 \cup X_2 = \{1,2,\ldots,n\}$. The basis of such bipartite graph is $X_1 = \{1,2,\ldots,k\}$, $0 \leq k \leq n$. If $U = (X_1, X_2, E(U))$ and $V = (Y_1, Y_2, E(V))$ then $U \leq V$ iff $|X_1| \leq |Y_1|$ and $\{(i,j) \in E(U): j > |Y_1|\} = \{(i,j) \in E(V): i \leq |X_1|\}$. This partial order is a natural partial order of subobjects of an object in a triangular category with bipartite graphs as morphisms.
**KEYWORDS:** Bipartite graph; partial order; triangular category.

## 1 The set $B_n$ of bipartite graphs

We restrict attention to finite simple graph and use standard notations and definitions of graph theory. A graph G is a pair (X, E), where X is a set $\{x_1, x_2, \cdots, x_n\}$ of elements called vertices, and E is a set of pairs of vertices $(x_i, x_j) = (x_j, x_i)$. An element $(x_i, x_j)$ of E is called an edge of G=(X, E). Any two vertices, $x_i$ and $x_j$, are said to be adjacent if and only if the pair $(x_i, x_j)$ is an edge of G. A graph (X, E) is bipartite if its vertices can be partitioned into two sets $X_1$ and $X_2$ ($X_1 \cup X_2 = X$; $X_1 \cap X_2 = \phi$) such that no two vertices in the same set are adjacent. One often writes U=($X_1$, $X_2$, E(U)) to denote a bipartite graph, and we say that the first set $X_1$ is the basis of the bipartite graph U.

An isomorphism of graphs G=(X, E) and G'=(X',E') is a bijection $f$: X $\to$ X' such that any two vertices $x_i, x_j \in$ X are adjacent in G if and only if $f(x_i), f(x_j) \in$ X' are adjacent in G'.

315

Now, we denote by $B_n$ the set of bipartite graphs $U=(X_1, X_2, E(U))$ with $n$ vertices such that the following laws hold:
1) The family $\{x_1, x_2, \cdots, x_n\}$ of the vertices of U is denoted by its set of indices $\{1,2,\cdots,n\}$ such that the first indices $(i = 1,2,\cdots,k)$ are in the same partite set namely in the basis $X_1$ of U and $X_2=\{k+1, k+2, \cdots, n\}$.
2) If $i$ and $j$ are adjacent in U such that $i \in X_1$ and $j \in X_2$ then $(i,j)$ denotes the corresponding edge of U. Thus $(i,j) \in E(U)$ implies $i < j$.

For instance, the following bipartite graph U:

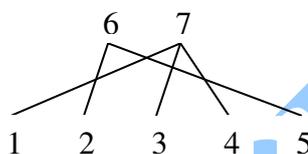

with the set of edges E={(1,7),(2,6),(4,7),(5,6)} and $X_1$={1,2,3,4,5}, $X_2$={6,7} is an element of $B_7$. The following two elements of $B_7$:

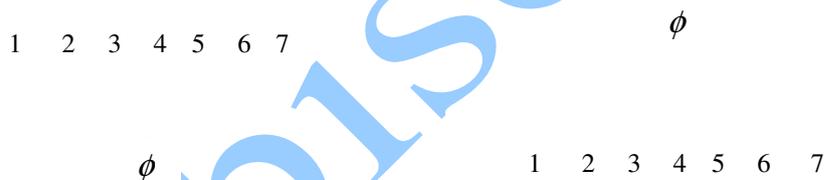

are two distinct elements of $B_7$. The first has basis the empty set, and {1,2,3,4,5,6,7} is the basis of the second bipartite graph.
It is straightforward to check that

$$|B_n| = \sum_{k=0}^{n} 2^{k(n-k)},$$

where $|B_n|$ is the cardinality of $B_n$.

## 2. A partial order on $B_n$

Let $U = (X_1, X_2, E(U))$ and $V = (Y_1, Y_2, E(V))$ be two elements of $B_n$.

**Proposition 1.** *The relation defined by*
$U \leq V \iff |X_1| \leq |Y_1|$ and $\{(i,j) \in E(U) \mid j > |Y_1|\} = \{(i,j) \in E(V) \mid i \leq |X_1|\}$
*is a partial order on $B_n$.*

316

**Proof.** It is immediate that the relation defined above is reflexive and anti-symmetric. To show that it is also transitive, let $U = (X_1, X_2, E(U))$, $V = (Y_1, Y_2, E(V))$, $W = (Z_1, Z_2, E(W)) \in B_n$ such that $U \leq V$ and $V \leq W$. It follows that

a)
$$|X_1| \leq |Y_1| \leq |Z_1|,$$

b)
$$(i_0, j_0) \in \{(i, j) \in E(U) \mid j > |Z_1|\} \Rightarrow$$
$$\Rightarrow (i_0, j_0) \in \{(i, j) \in \{(i, j) \in E(U) \mid j > |Y_1|\} = \{(i, j) \in E(V) \mid i \leq |X_1|\}$$
$$\Rightarrow (i_0, j_0) \in \{(i, j) \in E(V) \mid i \leq |X_1| \text{ and } j > |Z_1|\} = \{(i, j) \in E(W) \mid i \leq |X_1|\}$$
$$\Rightarrow \{(i, j) \in E(U) \mid j > |Z_1|\} \subseteq \{(i, j) \in \{(i, j) \in E(W) \mid i \leq |X_1|\}$$

c)
$$(i_0, j_0) \in \{(i, j) \in E(W) \mid i \leq |X_1|\} \Rightarrow$$
$$\Rightarrow (i_0, j_0) \in \{(i, j) \in E(W) \mid i \leq |Y_1|\} = \{(i, j) \in E(V) \mid j > |Z_1|\}$$
$$\Rightarrow (i_0, j_0) \in \{(i, j) \in E(V) \mid i \leq |X_1| \text{ and } j > |Z_1|\} = \{(i, j) \in E(U) \mid j > |Z_1|\}$$
$$\Rightarrow \{(i, j) \in E(W) \mid i \leq |X_1|\} \subseteq \{(i, j) \in E(U) \mid j > |Z_1|\}$$

a), b) and c) implies that $U \leq W$.

**Proposition 2.** *If the bases of two elements $U, V \in B_n$, $U \neq V$, are equal then $U$ and $V$ are incomparable.*

**Proof.** The equality
$$\{(i, j) \in E(U) \mid j > |Y_1|\} = \{(i, j) \in E(V) \mid i \leq |X_1|\}$$
where $X_1 = Y_1$, implies that $U = V$.

Now, let $n=3$. Then the Hasse diagram of the partially ordered set $(B_3, \leq)$ is the following one:

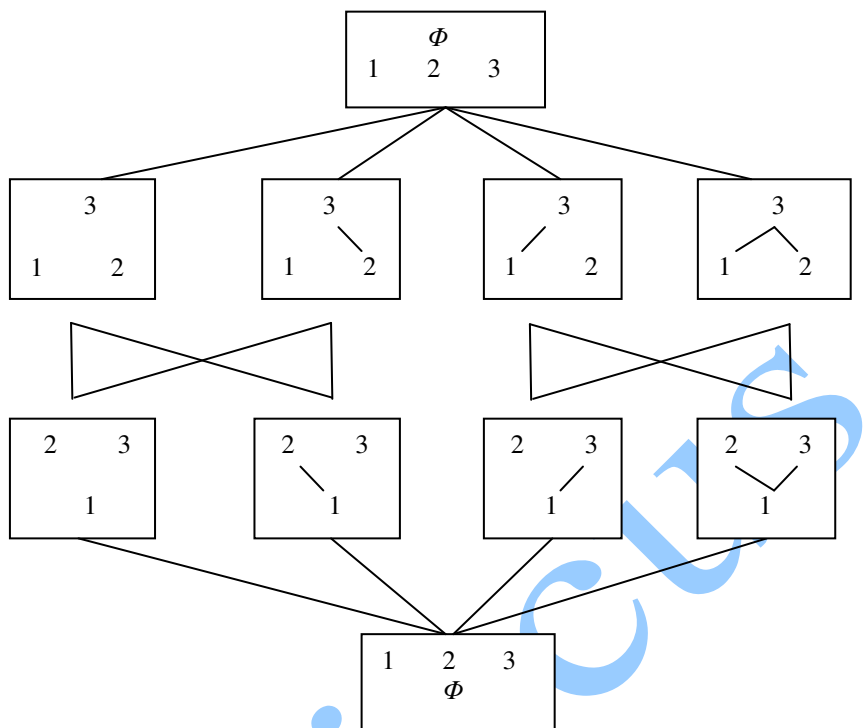

Without specifying the bipartite graphs, the Hasse diagram of $(B_3, \leq)$ is given by:

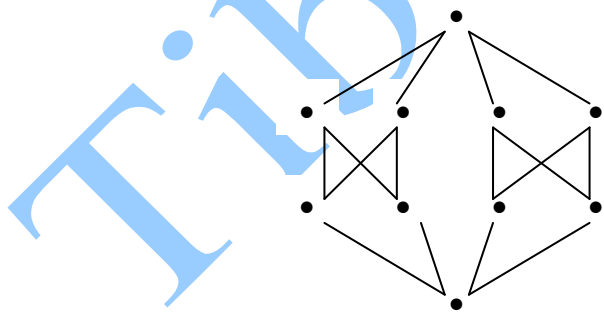

The case $n > 3$ is somewhat laborious. For example, the Hasse diagram of the partial ordered set $(B_4, \leq)$ is the following one:

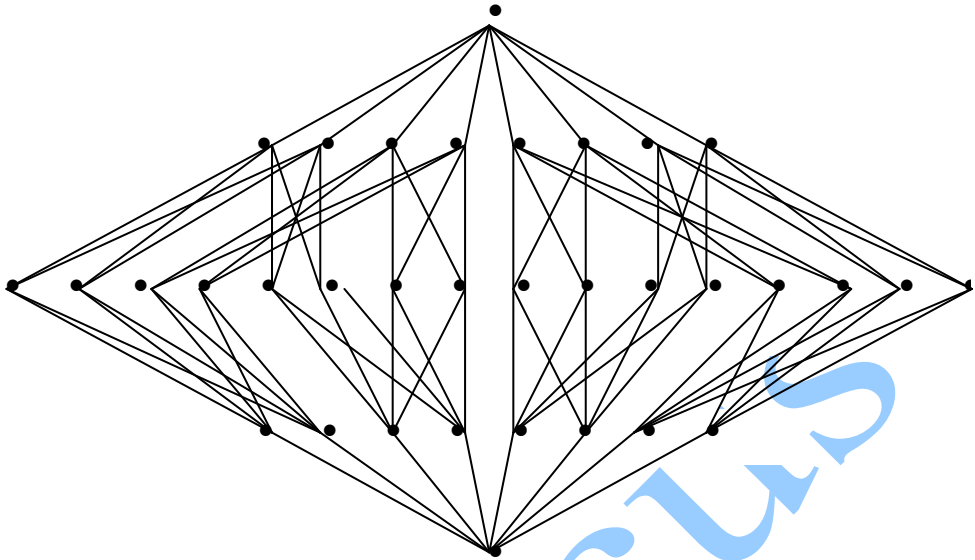

## 3. Connection with a triangular category

Möbius inversion for categories was considered for the first time by Leroux [Ler75]. A Möbius category in the sense of Leroux is a decomposition finite category C (i.e. a small category where each morphism $\alpha$ has only finitely many nontrivial factorizations) such that an incidence function $f : MorC \to$ has a convolution inverse if and only if $f(1_A) \neq 0$ for any identity morphism $1_A$ of C.

The convolution $f * g$ of two incidence functions $f$ and $g$ is defined by

$$(f * g)(\alpha) = \sum_{\alpha = \beta\gamma} f(\beta) \cdot g(\gamma) \qquad (\alpha \in MorC).$$

Möbius categories have also been characterized as decomposition-finite categories in which

(1) each identity morphism is indecomposable, i.e., $1_A = \beta\gamma$ implies $\beta = 1_A = \gamma$;
(2) $\beta\gamma = \gamma$ implies that $\beta$ is identity morphism.

Now, it is straightforward to see that a special class of categories (called triangular categories by Leroux [Ler80]) in which the set of objects is the set of nonnegative integers N and the family of numbers |*Hom*(*k,n*)|

319

(where |*Hom*(*k*,*n*)| is the number of morphisms from *k* to *n*) constitute a triangular family of numbers, that is:

$|Hom(n,n)| = 1$ *for all* $n \in N$; *and* $|Hom(m,n)| = 0$ *if* $m > n$.

The prime example of a triangular category (denoted $\Delta$ in [Ler80]) is that for which $0 \in N$ is the initial object and $\text{Hom}_\Delta(k,n)=$"the set of all injective and isotone maps from $\{1,2,…,k\}$ to $\{1,2,…,n\}$". The corresponding triangular family of numbers is the following one:

$$|Hom_\Delta(k,n)| = \binom{n}{k} \qquad (k \leq n)$$

More combinatorial triangular families of numbers can be represented by triangular categories (see [Ler80], [Ler90]).

Let C be a triangular category. The set S(*n*) of subobjects of *n*∈N (or, rather, monomorphisms into *n*) is an ordered set:

$$\alpha \leq \beta \quad \Leftrightarrow \quad \exists \gamma : \beta\gamma = \alpha$$

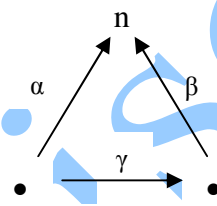

This relation is called the natural partial order on the set of subobjects of *n*.

**Proposition 3.** *Since* $|Hom(k,k)| = 1$ *for every* $k \in N$, *two monomorphisms* $\alpha, \beta$ *into n*, $\alpha \neq \beta$, *with the same domain k are incomparable*.

A triangular category is called lattice-triangular if $(S(n), \leq)$ is a lattice for every *n*∈N. A triangular category C is called monomorphic-triangular if any morphism of C is a monomorphism. We have:

**Theorem 4.** ([Sch03]) *Let* C *be a monomorphic-triangular category. Then* C *is lattice- triangular if and only if* C *has pullbacks.*

The triangular category $\Delta$ is a lattice triangular category. It is straightforward to check that in $\Delta$ the lattice $(S(n), \leq)$ is isomorphic to the Boolean algebra of all subsets of the set $\{1,2,…,n\}$. This category is not a category with pushout and therefore in Theorem 1, the "pullback" cannot be replaced by "pushout".

Now, we shall consider the category *B* of bipartite graphs (see *Bipis* in [Ler80]) defined by:

320

- Ob$B$=N;
- $Hom_B(k,n) = \begin{cases} \{U \in B_n \mid \{1,2,\cdots,k\} \text{ is the basis of } U\} & \text{if } k \leq n \\ \phi & \text{if } k > n \end{cases}$
- The composition of two morphisms: if $U \in Hom_B(m,k)$ and $V \in Hom_B(k,n)$ then the composition $V \bullet U$ is the bipartite graph with *n* vertices, 1,2,…,*n*, having the basis {1,2,…,*m*}; and the set of edges being the union of the set of edges of U and the set of those edges of V which have an endpoint in the basis of U.

For example, if $U \in Hom_B(2,5)$ is given by

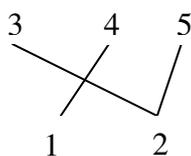

and if $V \in Hom_B(5,7)$ is given by

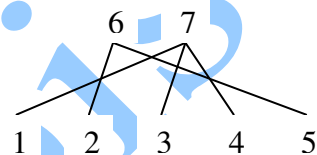

then

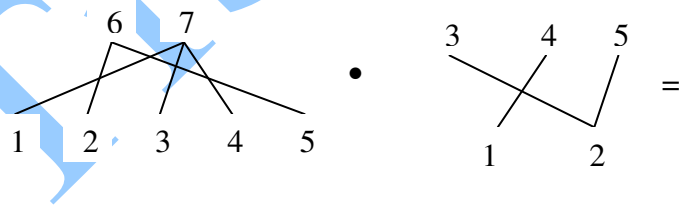

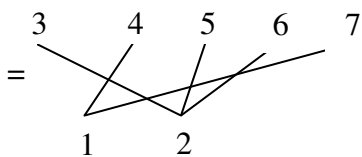

that is,

$$E(V \bullet U) = \{(1,4);(2,3);(2,5);(1,7);(2,6)\}.$$

Hence,
$$E(V \bullet U) = E(U) \cup \{(i,j) \in E(V) \mid i \leq m\}.$$
The identity morphism from *n* to *n* is the bipartite graph with *n* vertices and with basis $\{1,2,\ldots,n\}$; the set of edges being the empty set.

**Proposition 5.** *The category B is a triangular category and the corresponding triangular family of numbers is the following one:*
$$\mid Hom_B(k,n) \mid = 2^{k(n-k)} \qquad (k \leq n)$$

**Proposition 6.** *The category* B *is monomorphic but it is not epimorphic.*

**Proof.** Let $V \in Hom_B(m,n)$ and $U, U' \in Hom_B(k,m)$ be such that
$$V \bullet U = V \bullet U'.$$
Then,
$$U \cup \{(i,j) \in V \mid i \leq k\} = U' \cup \{(i,j) \in V \mid i \leq k\}$$
and therefore $U = U'$.

Now, if $V' \in Hom_B(m,n)$ such that
$$V \bullet U = V' \bullet U$$
then,
$$\{(i,j) \in V \mid i \leq k\} = \{(i,j) \in V' \mid i \leq k\}.$$
But this does not imply that $V = V'$.

**Proposition 7.** *The partial order on $B_n$ is the natural partial order on the set of subobjects of n in the triangular category B.*

**Proof.** Let $U = (X_1, X_2, E(U))$ and $V = (Y_1, Y_2, E(V))$ be two elements of $B_n$ such that
$$\mid X_1 \mid \leq \mid Y_1 \mid \text{ and } \{(i,j) \in E(U) \mid j > \mid Y_1 \mid\} = \{(i,j) \in E(V) \mid i \leq \mid X_1 \mid\}.$$
These two bipartite graphs *U* and *V* are two morphisms of the category *B* having the same codomain *n*. Consider the morphism $W = (X_1, Y_1 - X_1, E(W)) \in Hom_B(\mid X_1 \mid, \mid Y_1 \mid)$, where
$$E(W) = \{(i,j) \in E(U) : j \leq \mid Y_1 \mid\}$$
and we obtain:
$$E(V \bullet W) = E(W) \cup \{(i,j) \in E(V) : i \leq \mid X_1 \mid\} =$$
$$= \{(i,j) \in E(U) : j \leq \mid Y_1 \mid\} \cup \{(i,j) \in E(U) : j > \mid Y_1 \mid\} = E(U).$$
It follows that the diagram

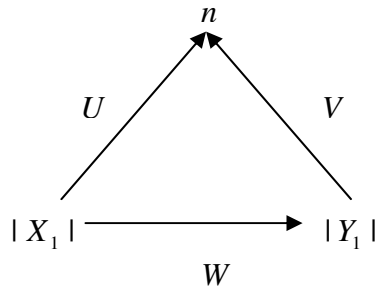

is commutative in *B*.

Conversely, if the above diagram is commutative in *B*, where $U = (X_1, X_2, E(U))$, $V = (Y_1, Y_2, E(V))$ and $W = (X_1, Y_1 - X_1, E(W))$, then
$$|X_1| \leq |Y_1|$$
and
$$\left. \begin{array}{l} E(W) = \{(i,j) \in E(U) : j \leq |Y_1|\} \\ E(U) = E(V \bullet W) = E(W) \cup \{(i,j) \in E(V) : i \leq |X_1|\} \end{array} \right\} \Rightarrow$$
$$\Rightarrow E(U) = \{(i,j) \in E(U) : j \leq |Y_1|\} \cup \{(i,j) \in E(V) : i \leq |X_1|\}$$
$$\Rightarrow \{(i,j) \in E(U) : j > |Y_1|\} = \{(i,j) \in E(V) : i \leq |X_1|\}$$

as required.

Taking into account the Hasse diagram of $(B_3, \leq)$ and Proposition 7, it follows:

**Proposition 8.** *The monomorhic-triangular category B is not a lattice-triangular category.*